\documentclass[fleqn,usenatbib]{mnras}

\usepackage[T1]{fontenc}

\usepackage{graphicx}	
\usepackage{amsmath}	
\usepackage{amssymb}	
\usepackage[normalem]{ulem}
\usepackage{tablefootnote}
\usepackage{pdflscape}
\usepackage{lscape}
\usepackage{gensymb}

\DeclareRobustCommand{\VAN}[3]{#2}
\let\VANthebibliography\thebibliography
\def\thebibliography{\DeclareRobustCommand{\VAN}[3]{##3}\VANthebibliography}

\title[Investigating the propagation of small-scale flare energy]{Investigating the propagation of small-scale flare  energy in the lower and upper atmosphere of solar active region}

\author[Gupta et al.]{Girjesh Gupta$^{1}$\thanks{E-mail: girjesh@prl.res.in (GG)}, Ananya Rawat$^{1,2}$, Helen Mason$^{3}$, Robertus Erd\'elyi$^{4,5,6}$ \\
$^{1}$ Udaipur Solar Observatory, Physical Research Laboratory, Dewali, Badi Road, Udaipur 313001, India \\
$^{2}$ Department of Physics, Indian Institute of Technology Gandhinagar, Palaj, Gandhinagar 382355, India \\
$^{3}$ DAMTP, Centre for Mathematical Sciences, University of Cambridge, Wilberforce Road, Cambridge CB3 0WA, UK \\
$^{4}$ School of Mathematical and Physical Sciences, University of Sheffield, Hounsfield Road, Sheffield S3 7RH, UK \\
$^{5}$ Department of Astronomy, E\"{o}tv\"{o}s Lor\'and University, P\'azm\'any P\'eter s\'et\'any 1/A, H-1117 Budapest, Hungary \\
$^{6}$ Gyula Bay Zolt\'an Solar Observatory (GSO), Hungarian Solar Physics Foundation (HSPF), Pet\H{o}fi t\'er 3., Gyula, H-5700, Hungary \\
}

\date{Accepted . Received ; in original form}

\pubyear{2025}

\begin{document}
\label{firstpage}
\pagerange{\pageref{firstpage}--\pageref{lastpage}}
\maketitle

\begin{abstract}

During solar flares, a considerable portion of the flare atmosphere becomes heated; however, the energy deposition process is still unclear, especially in the lower solar atmosphere. Here, we present spectroscopic and imaging observations of a small-scale transient of lifetime $<$1-min and further formation of a hot loop of lifetime $\approx$2-min in a solar active region. The observed transient shows the appearance of hot plasma $>$10 MK at the loop foot-point and the subsequent formation of a small-scale transient loop with a loop-top temperature $>$8 MK. The transient shows an enhancement in intensities in several AIA and IRIS passbands. Light curves obtained from several lower atmospheric passbands show consistent time lags in several peak intensities, which, to our knowledge, has never been reported before. Beneath the transient, associated HMI magnetogram shows evidence of flux emergence of both polarities. Using the IRIS \ion{O}{IV} line pair, we obtained the average electron number density of $10^{11.22}$ cm$^{-3}$ at the transient. IRIS transition region lines such as \ion{O}{IV} and \ion{Si}{IV} show a redshift of 10-15 km s$^{-1}$, whereas neutral lines such as \ion{C}{I} and \ion{S}{I} show a redshift of about 5 km s$^{-1}$. These Doppler shifts suggest a down-flowing warmer plasma in the lower atmosphere. The appearance of \ion{Mg}{II} triplets in emission is also observed.  We interpret these enhancements in intensities in the lower atmosphere as a result of heating due to both non-thermal electrons and thermal conduction operating simultaneously.

\end{abstract}




\section{Introduction}
\label{sec:intro} 

The atmosphere of the Sun is highly dynamic and diverse, and hosts several small-scale events and structures \citep[e.g.,][]{2014Sci...346C.315P,2014Sci...346E.315H}. These small-scale transients and brightenings may contribute to the mass and energy requirements of the upper atmosphere \citep[e.g.,][]{2015ApJ...809...82G}. Reports on several low-lying small-scale loops at very high temperatures, which evolve on time scales of a few minutes, also exist in the solar atmosphere \citep[e.g.,][]{2014Sci...346B.315T,2018ApJ...857..137G,2018A&A...615L...9C}. These loops show transient activity on several spatial and temporal scales. This activity is generally associated with magnetic flux cancellation events at the photosphere \citep[e.g.,][]{2015ApJ...810...46H,2020A&A...644A.130C,2022MNRAS.512.3149G}.

During the impulsive energy release in flares and micro-flares, plasma in the chromosphere gets heated up and attains a temperature of more than $10$ MK. This enhanced temperature increases pressure at the chromosphere and pushes local plasma upward into the corona. This leads to the formation of post-flare loops. Such a process of loop filling with heated material is termed as 'chromospheric evaporation'  \citep[e.g.,][]{2009ApJ...699..968M,2018ApJ...857..137G}. On the other hand, high pressure also pushes denser plasma downward into the lower atmosphere, which is referred to as 'chromospheric condensation' \citep[e.g.,][]{2006ApJ...638L.117M}. In such models, emission from hot material from the corona is expected to show upward flows, and emission from cool material from the transition region and upper chromosphere is expected to show downward flows \citep[e.g.,][]{2017ApJ...841L...9L,2018ApJ...857..137G}. Since the density of chromosphere is much higher than the corona, the up-flow speed is much larger than the down-flow speed \citep[e.g.,][]{1985ApJ...289..414F,2009ApJ...699..968M,2018ApJ...857..137G}.

Observations of foot-point brightening associated with coronal micro/nano-flares have provided indirect evidence of the presence of non-thermal electrons in the small heating events \citep[e.g.,][]{2014Sci...346B.315T}. Upon combining the Interface Region Imaging Spectrograph \citep[IRIS;][]{2014SoPh..289.2733D} and simulation results, some of the spectral properties of the IRIS transition region and chromospheric lines depend on the mechanism of energy transport, i.e., thermal conduction or non-thermal particles, and also on the time duration of heating \citep[e.g.,][]{2023FrASS..1014901P}. These simulation results have shown that blueshift of IRIS \ion{Si}{IV} lines and emission in IRIS \ion{Mg}{II} triplets are signatures of heating due to non-thermal particles, whereas redshift of IRIS \ion{Si}{IV} lines are signatures of heating due to thermal conduction \citep[e.g.,][]{2014Sci...346B.315T,2018ApJ...856..178P,2020ApJ...889..124T}. In recent times, growing evidence has emerged of the direct presence of non-thermal electrons in very small transient events \citep[sub-A class flares;][]{2021MNRAS.507.3936C}, and a wealth of evidence in the larger microflares \citep[A and B class flares;][]{2024A&A...691A.172B}.

In this paper, we present a detailed investigation of the upper and lower solar atmospheric response of a small-scale heating event at the loop foot-point using multi-wavelength imaging and spectroscopic data. In Section~\ref{sec:obs}, we present the multi-wavelength observations of a small-scale transient and the associated loop. Imaging, spectroscopic, and magnetic properties of the event are described in Section~\ref{sec:analysis}. Finally, the discussion and conclusion of our findings are provided in Section~\ref{sec:disc}.

\section{Observations}
\label{sec:obs}

\begin{figure}
   \centering
   \includegraphics[width=0.5\textwidth]{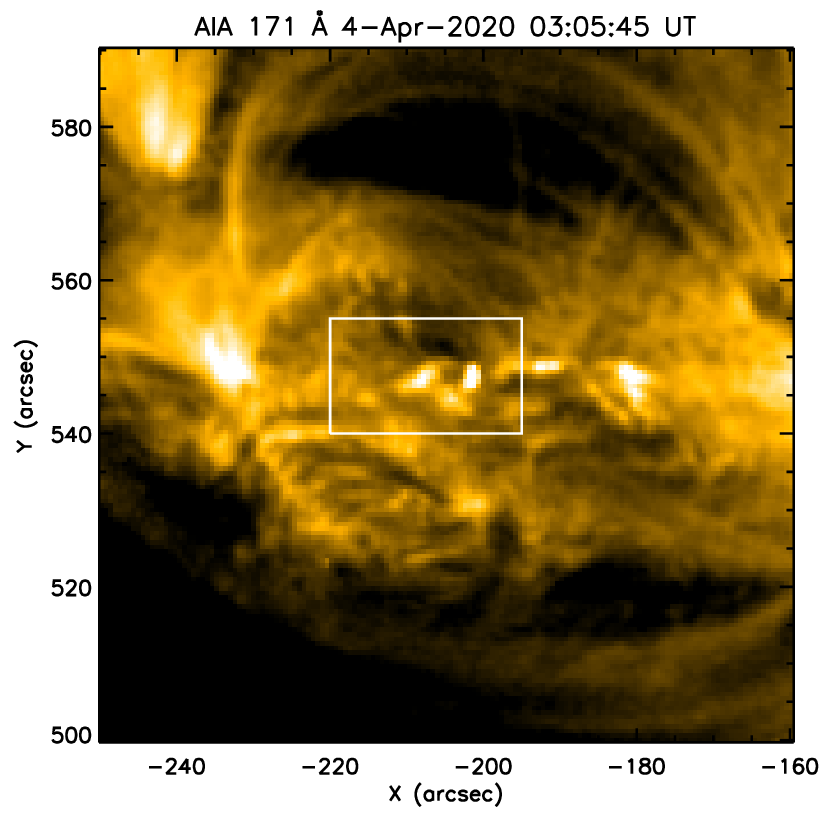}
   \caption{Image of active region AR 12759 as recorded by AIA 171~\AA\ passband on 4 April 2020 chosen for our analysis. The white box shows the region selected for detailed investigation.}
    \label{fig:context}
\end{figure}

\begin{figure*}
    \centering
    \includegraphics[width=\textwidth]{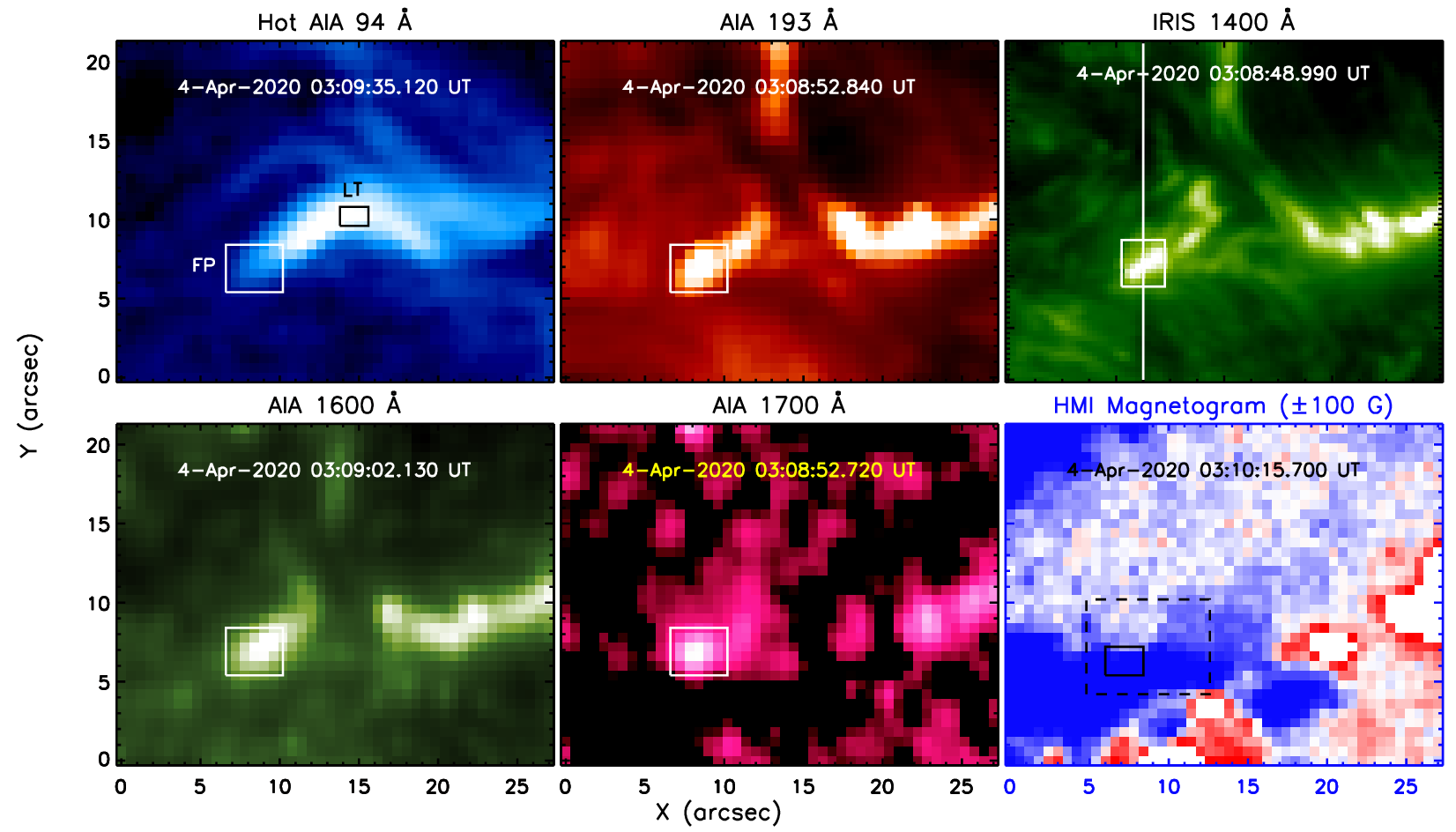}
    \caption{Images obtained from different AIA, HMI, and IRIS-SJI passbands as labelled, showing the region of detailed investigation. The HMI magnetic field image is scaled between $\pm$100~G.  Different boxes on the AIA, HMI, and IRIS images represent the location of the transient, which is one of the foot-points (FP) of the transient loop under study and associated loop-top (LT). The vertical white line on IRIS 1400 \AA\ image represents the position of the IRIS slit at the time as labelled. The image from AIA 1700 \AA\ is a difference image with respect to an image obtained around 3:07 UT. }
    \label{fig:contextall}
\end{figure*}

To investigate the origin and characteristics of a small-scale transient, associated heating of the lower atmosphere and subsequent formation of the hot transient loop in the solar active region, we identified a suitable dataset observed by the Atmospheric Imaging Assembly \citep[AIA;][]{2012SoPh..275...17L}, the Helioseismic and Magnetic Imager \citep[HMI;][]{2012SoPh..275..207S,2012SoPh..275..229S}, and the Interface Region Imaging Spectrograph \citep[IRIS;][]{2014SoPh..289.2733D}. AIA and HMI are both onboard the Solar Dynamics Observatory \citep[SDO;][]{2012SoPh..275....3P}. The IRIS dataset provides good high-resolution coverage over the lower atmosphere, while the IRIS spectroscopic slit provides spectral data for density diagnostics of the transition region. We also utilize X-ray data from the X-ray Telescope \citep[XRT;][]{2007SoPh..243...63G} onboard Hinode and the Solar X-ray Monitor \citep[XSM;][]{2014AdSpR..54.2021V} onboard Chandrayaan-2 to complement our observations.

AIA onboard SDO provides full-disk images of the Sun in seven EUV passbands that manifest the upper solar atmosphere and three UV-visible passbands that manifest the lower solar atmosphere. HMI onboard SDO provides full disk photospheric images of the Sun in intensity continuum, Dopplergram, and magnetogram. These observables are extracted using \ion{Fe}{I} 6173 \AA\ line. HMI magnetogram data are utilized to explore the magnetic origin of the transient in the lower solar atmosphere. These instruments have been continuously observing the Sun since their launch in 2010. All the images were calibrated, co-aligned, and re-scaled to a common 0.6$\arcsec$/pixel resolution using the standard $aia\_prep.pro$ and $hmi\_prep.pro$ routines available in the standard Solar Software \citep[SSW;][]{1998SoPh..182..497F}.

IRIS performed two 320-step raster of the region on 04 April 2020 from 02:11:53 to 03:50:04 UT using 175'' long slit in a step size of $\approx 0.35''$. Spectral slit has pixel resolution 0.33''. Thus, the rasters covered the field-of-view (FOV) of approximate $112'' \times 175''$. Each slit position has an exposure time of $\approx 8$ s and a cadence of $\approx 9.2$ s. Each raster took 48 minutes and 57 seconds to complete the scan. The Slit-Jaw Imager (SJI) of IRIS observed in 1330, 1400, 2796, and 2832 \AA\ passbands with an exposure time of $\approx 8$ s and an effective cadence of $\approx 36.8$ s. IRIS-SJI covered the  FOV of approximate $167'' \times 175''$. In this work, we utilize IRIS level-2 data. Slit-jaw images from all the passbands are co-aligned in the IRIS level-2 data distribution.

Imaging data enables us to co-align images from multiple instruments, thereby allowing us to utilize data obtained from  instruments operating at various wavelengths. We aligned the IRIS data with respect to AIA data using images from the IRIS-SJI 1400 \AA\ and AIA 1600 \AA\ passbands. We also co-aligned HMI data with aligned IRIS data using IRIS-SJI 2832 \AA\ and HMI continuum images. All the alignments were carried out employing the SSW routine $align\_map.pro$. All the passbands are aligned within the accuracy of 1'' with each other.  All images from IRIS, AIA, and HMI instruments are further de-rotated to the time 02:50 UT using the SolarSoft programs. The selected dataset offers a rare opportunity to explore the temporal evolution of a small-scale transient throughout the whole solar atmosphere.

Figure~\ref{fig:context} shows part of active region 12759 selected for our analysis. The panel shows an image recorded in the AIA 171  \AA\ passband before the transient. The over-plotted white box shows the region selected for a detailed investigation.  In Figure~\ref{fig:contextall}, we show images of the transient and associated newly formed coronal loop at the time of the peak of its intensity evolution in the different passbands as labelled. Different images probe different temperatures of the solar atmosphere, as various AIA, HMI, and IRIS-SJI passbands are sensitive to different plasma temperatures. The IRIS-SJI 1400 \AA\ passband is sensitive to a temperature of about 80 kK \citep{2014SoPh..289.2733D}, AIA 94, and 193 \AA\ passbands are sensitive to 7 MK, and 1.5 MK plasma temperatures, respectively \citep{2010A&A...521A..21O,2012SoPh..275...17L}, whereas HMI magnetogram data are taken at the solar photosphere \citep{2012SoPh..275..229S}. 

In Figure~\ref{fig:contextall}, the hot transient loop is clearly visible in the AIA 94 \AA\ passband (sensitive to $\approx$7 MK plasma), whereas the AIA 193 \AA\ passband shows the peak intensity phase of the transient, which led to the formation of a hot loop observed in the AIA 94 \AA\ passband. Down in the lower atmosphere, transient signatures were clearly observed in the IRIS 1400 \AA\  and AIA 1700 \AA\ passbands, though smaller in size. During the peak intensity phase of the transient observed in the IRIS-SJI 1400 \AA\ passband, the IRIS spectroscopic slit was passing over the transient. 
The HMI line-of-sight magnetogram shows a mixed polarity region underneath the transient.  Therefore, the identified dataset provides a rare opportunity to study the evolution of a small-scale event, heating of the lower atmosphere, and formation of an associated transient loop using imaging and spectroscopic data, along with the changes observed in the photospheric magnetic field beneath.

\begin{figure*}
    \centering
    \includegraphics[width=0.5\textwidth]{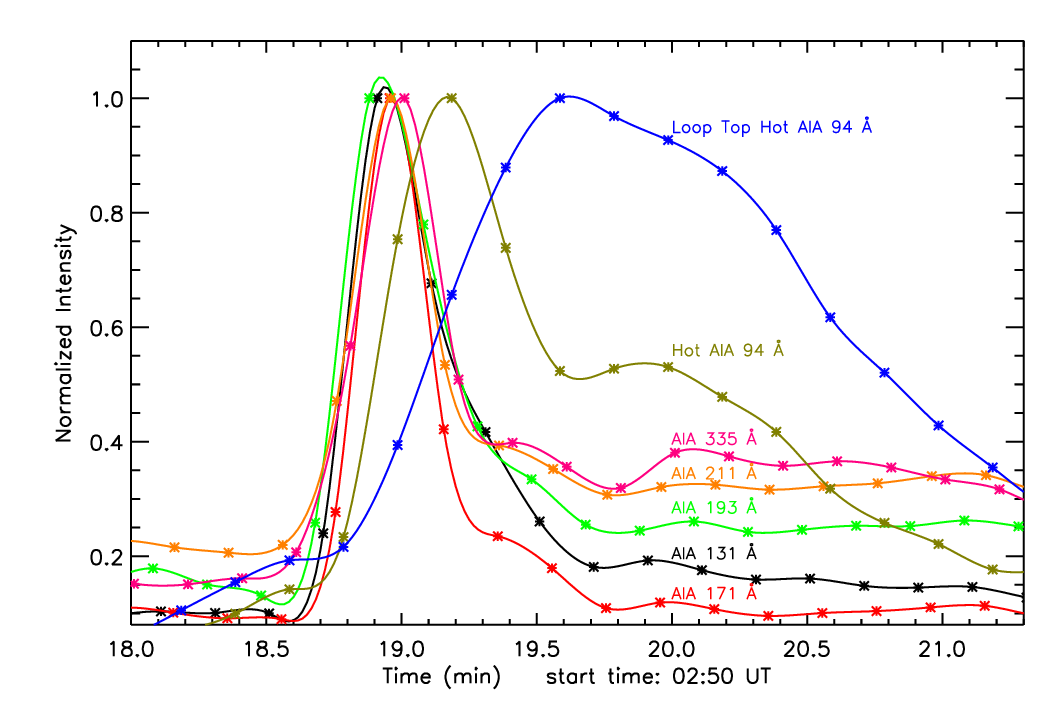}\includegraphics[width=0.5\textwidth]{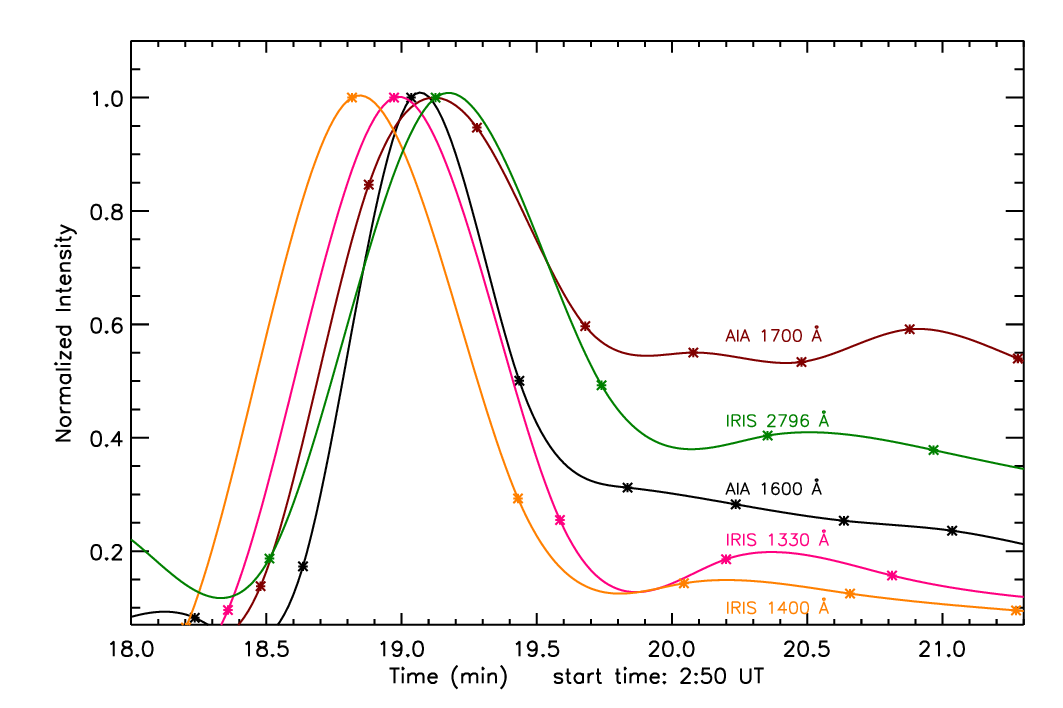}
    \caption{The intensity evolution obtained at the transient loop foot-point in the boxed region from various AIA and IRIS passbands, as labelled, sensitive to coronal emissions (left panel) and lower atmospheric emissions (right panel). Asterisk (*) symbols represent observed data points, whereas overplotted continuous lines are spline interpolated light curves.}
    \label{fig:lc}
\end{figure*}

\section{Data Analysis and Results}
\label{sec:analysis}

Here, we focus our attention on the small-scale event and subsequent appearance of a transient loop. Here, we present in greater detail the imaging, spectroscopic, and magnetic properties of the observed small-scale transient.

\subsection{UV-visible Imaging Analysis}
\label{sec:UVimaging}

Figure~\ref{fig:contextall} shows images of the transient foot-point and the associated newly formed coronal loop in different passbands as labelled. The transient loop is clearly seen in the AIA 94 \AA\ passband. The AIA 94 \AA\ passband is sensitive to hot plasma due to \ion{Fe}{XVIII} 93.93 \AA\ emission (peak formation temperature $\approx$7 MK); however, this passband suffers from strong contributions from cooler temperature emissions \citep[][]{2011A&A...535A..46D}. We have removed these cool emission components from the AIA 94 \AA\ passband using the approach developed by \citet{2013A&A...558A..73D} as follows 
\begin{equation}
 I(Fe~XVIII) \approx I(94~{\AA}) - I(211~{\AA})/120 - I(171~{\AA})/450
\end{equation}
where  I(94~{\AA}), I(211~{\AA}), and I(171~{\AA}) are intensities from AIA 94~{\AA}, 211~{\AA}, and 171~{\AA} passbands, respectively. Henceforth, in Figure~\ref{fig:contextall}, we show emission detected from the hot plasma in \ion{Fe}{XVIII} 93.93 \AA\ spectral line as hot AIA 94 \AA\ ($\approx$7 MK plasma). The panel clearly demonstrates the appearance of a complete transient hot coronal loop after the foot-point brightening observed in panels AIA 193, IRIS 1400, and AIA 1700 \AA\ as labelled. The transient foot-point brightening has a lifetime of about 1 minute and a length-scale of about 1300 km, whereas the associated loop has a lifetime of about 2 minutes and a loop length of about 9--11 Mm. 

\begin{figure*}
    \centering
     \includegraphics[width=0.85\textwidth]{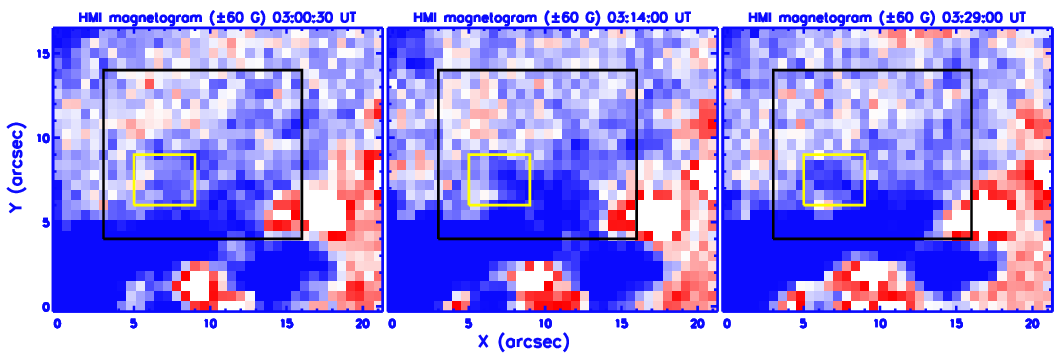}
    \includegraphics[width=0.85\textwidth]{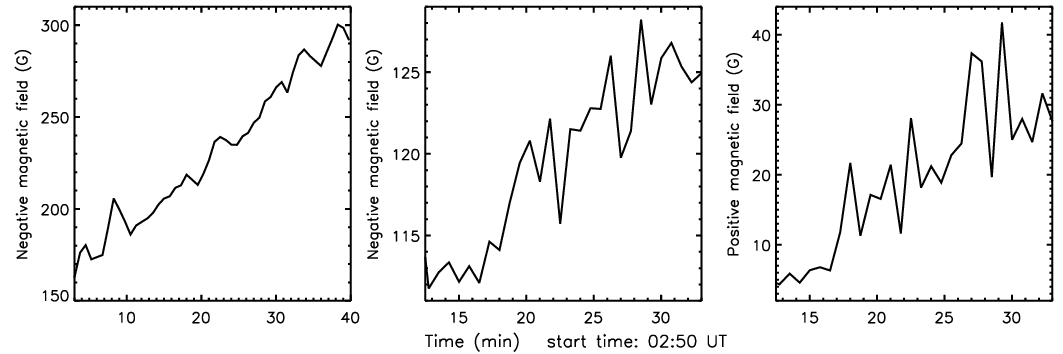}
    \caption{Upper panels: Images of the photospheric LOS magnetic field ($\pm$60 G) underneath the transient loop foot-point as recorded by HMI magnetogram. Lower panels: Variation of the magnetic field underneath the transient loop foot-point in the boxed regions marked on the HMI magnetogram images above. The left panel is for the smaller yellow box region, whereas the middle and right panels are for the bigger black box region.}
    \label{fig:mag}
\end{figure*}

Different boxes on the AIA, HMI, and IRIS-SJI images represent the location of the transient brightening under detailed investigation and are utilized to extract the average parameters in the different passbands.  The size of transient brightening decreases with decreasing temperature of the passbands. This is consistent with the expectation of magnetic flux tube expansion with height \citep[e.g.,][]{2023MNRAS.525.4815R,2025ApJ...988L..26R}, the size of the transient foot-point decreases into the lower atmosphere. The observed transient loop is termed a hot loop due to its appearance in the hot AIA 94 \AA\ passband, sensitive to approximately 7 MK plasma. 

\begin{figure*}
    \centering
    \includegraphics[width=0.95\textwidth]{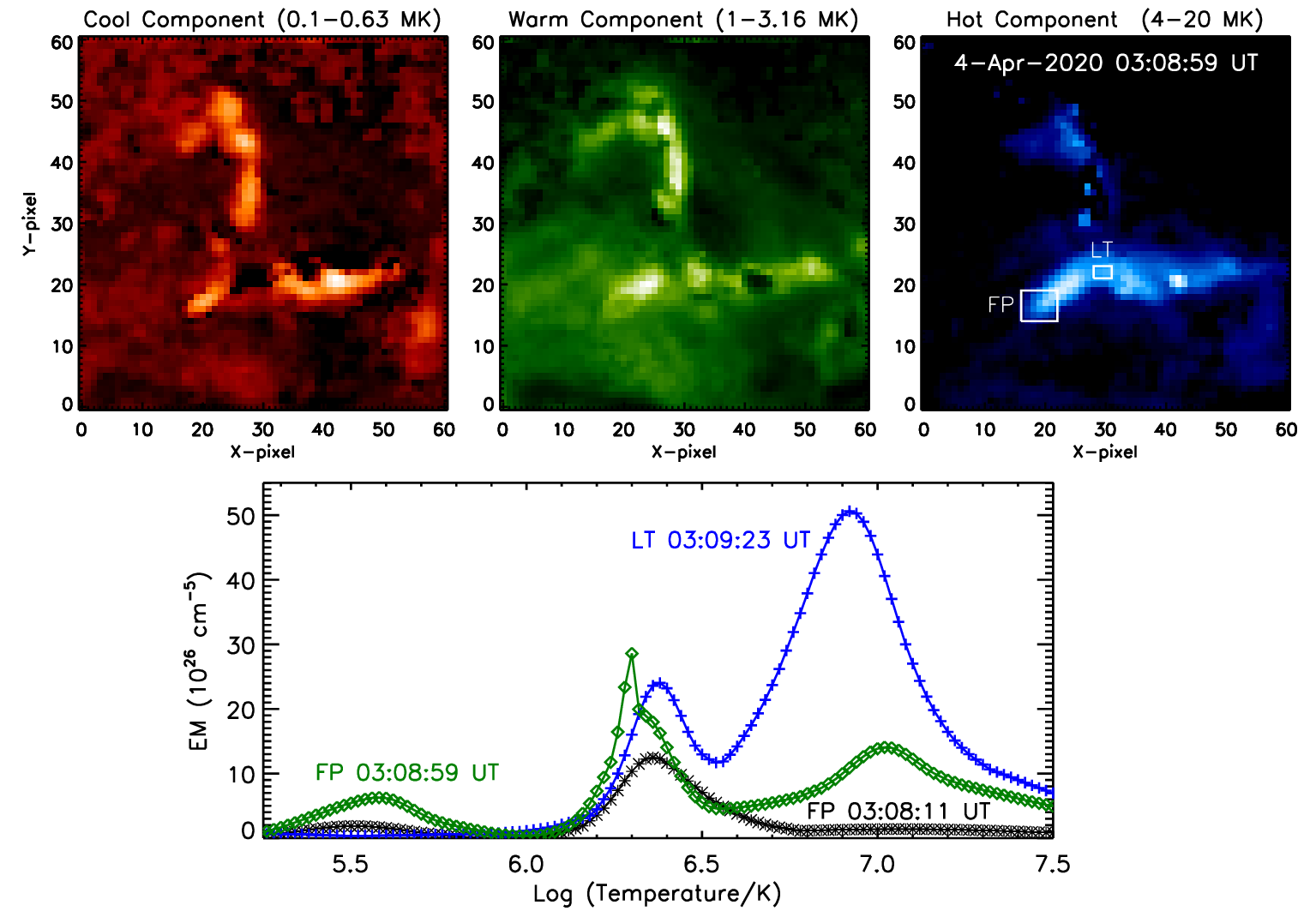}
    \caption{Differential emission measure (DEM) curve and images obtained at different plasma temperatures as labelled. DEM distribution curves are shown for different locations marked with FP and LT in DEM images at different times, as labelled.}
    \label{fig:dem}
\end{figure*}

In Figure~\ref{fig:lc}, we plot the evolution of the average intensity of the small-scale transient foot-point in the boxed region. The left panel shows the intensity evolution from the AIA passbands that are sensitive to coronal emissions, whereas the right panel shows intensity evolution from lower atmospheric emissions from the AIA and IRIS passbands, as labelled. Data points from AIA EUV passbands are obtained at an effective interval of 12 s, AIA UV passbands at 24 s, and IRIS passbands at 36.8 s. The observed data points are plotted with asterisk (*) symbols and are overplotted with the spline interpolated light curves obtained at every second. The intensity curves indicate that the transient region reached its maximum around 03:09 UT, but at different times in the coronal passbands such as AIA 335, 211, 131, 171, hot 94 \AA. However, we did not find any concrete pattern in time differences in different passbands with respect to their temperature sensitivity.  Interestingly, for lower atmospheric emissions, passbands such as AIA 1600, 1700 \AA, IRIS 1400, 1300, and 2796 \AA\ show some time-dependent intensity maxima, with the passbands sensitive to the hottest plasma peaks first, followed by cooler ones in order of decreasing temperature. The light curve from IRIS 1400 \AA\ passband peaks first ($\Delta t=0$ s) and is followed by  IRIS 1330 \AA\ passband ($\Delta t=8.0$ s); both have strong contributions from transition regions and subsequently in AIA 1600 and 1700 \AA\ passbands ($\Delta t=13$ s and $\Delta t=16$ s), which have significant contributions from temperature minimum region. 
This plot demonstrates the heating of the lower atmosphere, where heat flows from the upper atmospheric layer to the lower at the photosphere. Although chromospheric IRIS 2796 \AA\ passband peaks at $\Delta t=19$ s, the transient light curve from this passband is very broad compared to other passbands, and thus can have larger uncertainty in the peak time. Observations of the temperature-dependent evolution of the plasmas are commonly found at foot-points of flare loops \citep[e.g.][]{2013ApJ...774...14Q} and transient loops  \citep[e.g.][]{2022MNRAS.512.3149G}. Response of foot-point heating is first recorded from the lower atmosphere as noted by IRIS-SJIs 1400 and 1330 \AA\ and AIA 1600 \AA, and later at higher temperatures as recorded from AIA 171, 211, 335 \AA\ passbands. Therefore, this hot transient brightening can be considered as the miniature version of a standard flare and can be termed as micro-flare \citep[e.g.,][]{2018ApJ...857..137G}, here in this case, GOES A-class flare. Such a clear observation of time-dependent heating of the lower atmosphere of such a small class of solar flare is very interesting and has, to the best of our knowledge, never been reported before.  
Moreover, it was also noted by  \citet{2019ApJ...870..114S} that flare excess emissions observed in AIA 1600 and 1700 \AA\ passbands can be attributed to the chromosphere, where significant contributions from \ion{C}{I} 1561 and 1656 \AA\ multiplets, \ion{He}{II} 1640 \AA\ and other lines are present. In view of the above noted finding, it may also be possible that enhanced transient emissions observed in different lower atmospheric passbands may have their origin at the chromospheric height itself. Such a possibility can not be ruled out.

\begin{figure*}
    \centering
    \includegraphics[width=0.38\textwidth]{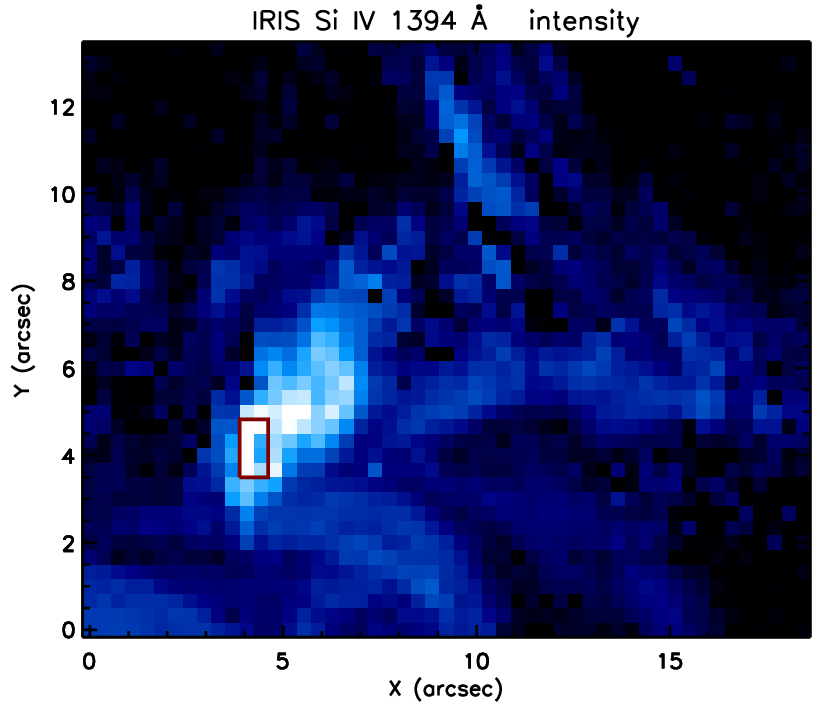}\includegraphics[width=0.62\textwidth]{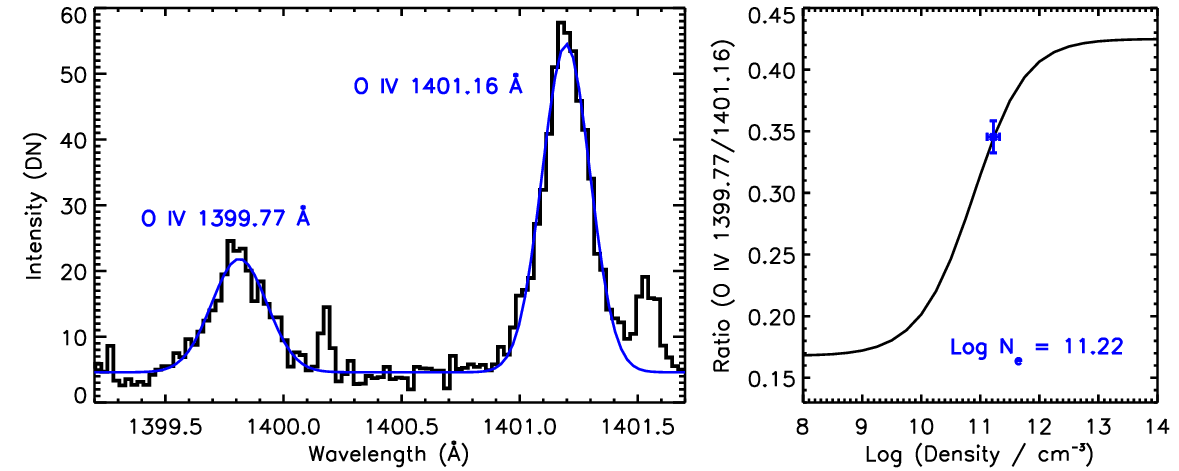}
    \caption{Left panel: IRIS spectroscopic raster images of the transient loop as obtained from the intensity of the \ion{Si}{IV} 1394 \AA\ line. The maroon box on the raster image represents the location of the transient loop foot-point under study. Middle panel: Average spectral line profile of the density-sensitive line pair of \ion{O}{IV} 1399 and 1401 \AA\ obtained at the transient loop foot-point (maroon box in left panel). Profiles are fitted with Gaussian functions, as shown in the blue line. Right panel: theoretically predicted intensity ratio of \ion{O}{IV} 1399 and 1401 \AA\ line pair with electron number density. The estimated intensity ratio and corresponding electron density at the transient loop foot-point are marked with an asterisk (*) and printed on the panel.}
    \label{fig:spec}
\end{figure*}

We further probed the driver of this small-scale transient using HMI LOS magnetic field data obtained at the photosphere. In Figure~\ref{fig:mag}, we show images of the evolution of the magnetic field beneath the transient region at the photosphere. The transient event is located almost above the mixed polarity region. We obtained average positive- and negative-polarity magnetic field evolutions beneath the transient event and within the black and yellow boxes in the top panels of Figure~\ref{fig:mag}. The total magnetic field increases with time in the box regions. In the smaller yellow box, the average magnetic field (of negative polarity) increases from 160 to 300 G over approximately 35 minutes. In the bigger black box, the average negative field strength increases from about 112 to 125 G, whereas the positive field increases from about 5 G to 30 G. We thoroughly investigated the field increase in this region with various box sizes and found that the observed increase in field strength is because of the emergence of magnetic flux and not due to any motion of magnetic patches into the rectangular boxes taken here. Furthermore, the smaller box contains no pixels with a positive-polarity magnetic field during the analysed time sequence. The increase in the magnetic field here with time can be considered as evidence of magnetic flux emergence.

\begin{figure*}
    \centering
    \includegraphics[width=0.75\textwidth]{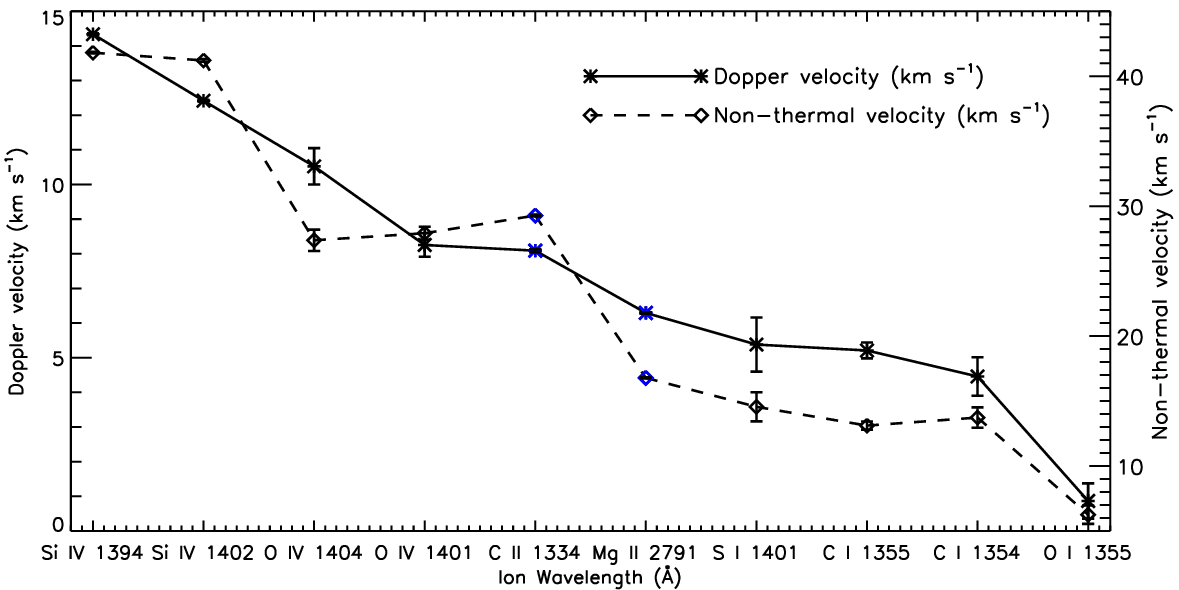}
    \caption{Doppler velocity and non-thermal velocity at the transient loop foot-point from the different IRIS spectral lines originating from different ions. Error bars are obtained from the fitting error on the wavelength and total width of the spectral profiles. \ion{C}{II} and \ion{Mg}{II} profiles are optically thick, so non-thermal velocities from these lines here provide upper limits. Estimates from these lines are plotted in blue.  Data points are arranged in almost decreasing order of their spectral line peak formation temperature. }
    \label{fig:velocity}
\end{figure*}

To quantify the temperature of the small-scale transient, we employ the differential emission measure (DEM) tool developed by \citet{2015ApJ...807..143C} that utilizes six EUV coronal passbands of AIA 94, 131, 171, 193, 211, and 335 \AA. In Figure~\ref{fig:dem}, we show emission measure (EM) images at cool (0.1-0.63 MK), warm (1-3.16 MK), and hot (4-20 MK) plasma components at 03:09 UT, i.e., at around the peak phase of the transient. Although loop foot-points are visible at all the plasma components, the transient loop is visible only at the hot component. We show the average EM curves at the loop foot-point and loop-top locations marked with boxes FP and LT, respectively, in the bottom panel.  We also plot the average EM curve in box FP, just before the transient, for comparison. Locations and times of EM curves are labelled accordingly. EM curve of the foot-point clearly shows the appearance of additional hot (6-14 MK) and cool (0.16 MK) plasma components during the transient with respect to before the transient time at the same transient foot-point location where plasma with a temperature of only 2.0-2.5 MK existed before. Therefore, the transient is composed of multi-thermal plasma around 8-12 (peaks at 10.5 MK), 1.5-2.5, and 0.1-0.6 MK. We also plot the EM curve at the loop-top during the peak of the transient, which shows enhanced hot plasma at about 8.5 MK, slightly cooler than the hot plasma component at the foot-point. 

We further calculated electron number densities at the transient using EM values at peak temperatures. For the purpose, we assume the plasma filling factor ($\phi$) to be 1 (i.e., $n_e^{euv}\approx \sqrt{EM/\phi l}$). At the peak of the transient, EMs at the foot-point are about  $14\times10^{26}$,  $29\times 10^{26}$,  and $6\times 10^{26}$ cm$^{-5}$ at 10.5, 2.0, and 0.4 MK, respectively. EUV brightening associated with the transient is observed over $l\approx 1300$ km (three AIA pixels wide). Hence, we estimated the densities to be about  $3.3\times10^9$, $4.7\times10^9$, and $2.1\times10^9$ cm$^{-3}$, respectively, for all three components of plasma. These densities are similar to those in the corona and represent the coronal response of the transient.

At the loop-top (LT), the EM curve peaks at around 8.5 MK temperature, and the corresponding peak EM is about $50\times 10^{26}$ cm$^{-5}$.  At this height, the loop width is $\approx1750$ km (four AIA pixels wide). Upon assuming the cylindrical cross-section of the loop and plasma filling factor ($\phi$) 1, we obtained the loop-top density to be $5.3\times10^9$ cm$^{-3}$, which is higher than the loop foot-point density at the hot plasma component. We note that all these densities derived from the EM using a filling factor of one are lower limits, since filamentary structures could be present. However, the plasma filling factor can vary along the coronal loop length between 0.1 and 1 at the coronal loop foot-point and loop-top, respectively \citep[e.g.,][]{2015ApJ...800..140G}. Therefore, upon assuming the plasma filling factor ($\phi$) 0.1 at the transient loop foot-point, the electron density comes out to be $\approx 1\times10^{10}$ cm$^{-3}$ for the hot plasma component, which is higher than the electron density obtained at the loop-top.

\subsection{UV Spectroscopic Analysis}
\label{sec:UVspect}

\begin{figure}
    \centering
    \includegraphics[width=0.5\textwidth]{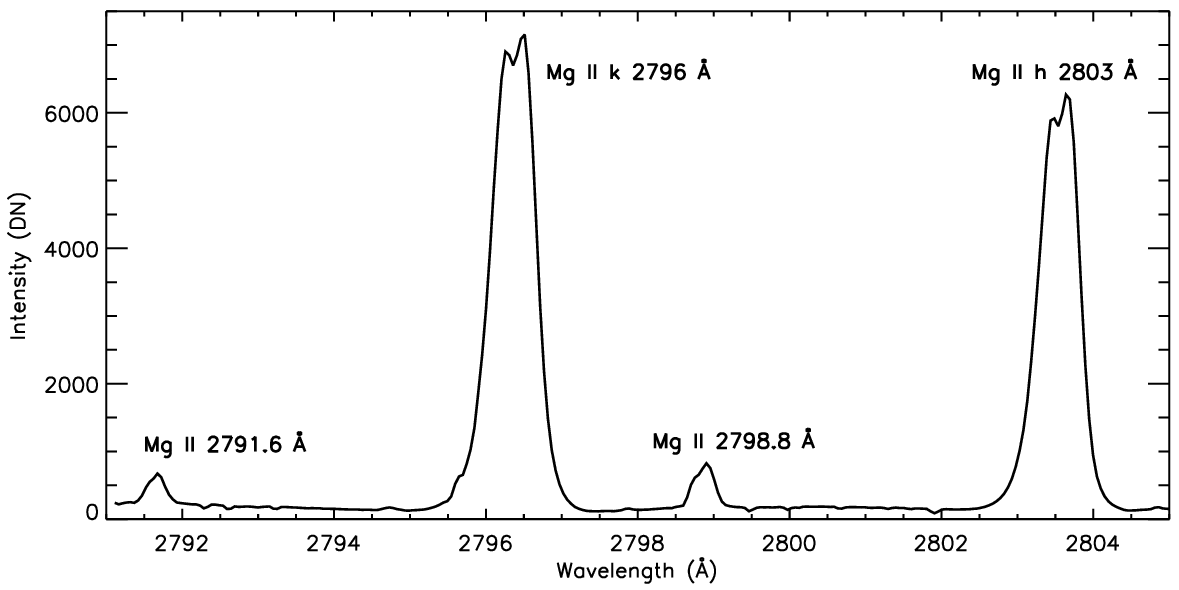}
    \caption{The appearance of \ion{Mg}{II} companion/subordinate triplet emission lines at wavelengths 2791.60, 2798.75, and 2798.82 \AA\  at transient loop foot-point.}
    \label{fig:mg2}
\end{figure}

\begin{figure*}
    \centering
    \includegraphics[width=1\textwidth]{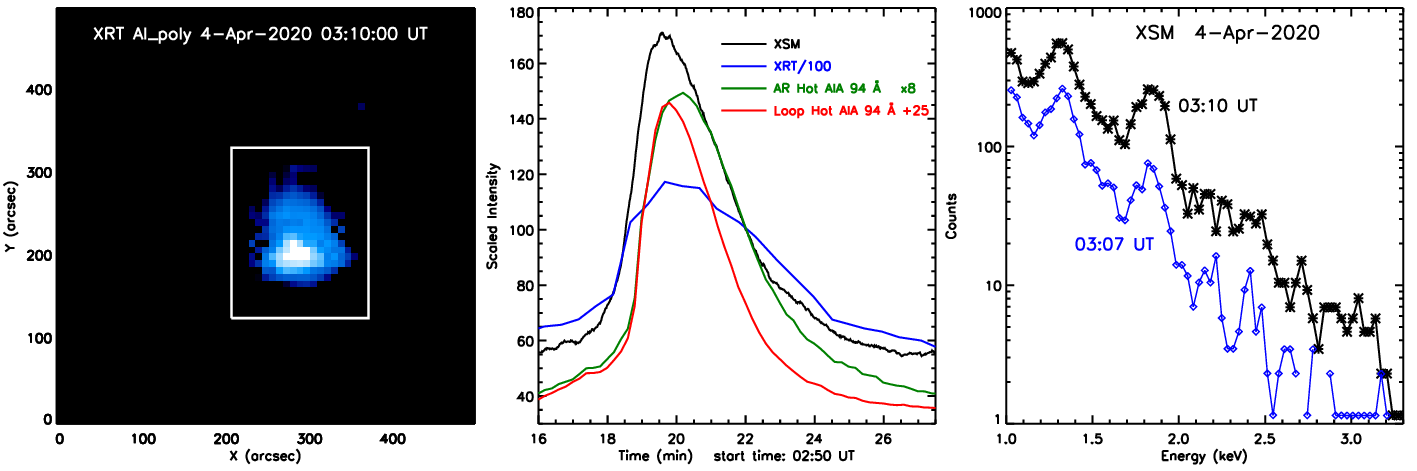}
    \caption{Left panel: XRT Al\_poly image of the active region AR 12759 at the time as labelled. The white box indicates the region chosen to extract the XRT light curve of the active region transient. Middle panel: XRT and Chandrayaan-2 XSM light curves showing analyzed A-class flare along with the hot AIA 94 \AA\ light curves for the whole active region and full loop. Intensities are scaled to fit in the plotting range. Right panel: Chandrayaan-2 XSM spectrum obtained at the peak phase of the analyzed A-class flare (03:10 UT) and before the transient (03:07 UT) as labelled.}
    \label{fig:xray}
\end{figure*}

Luckily, when the IRIS slit was rastering the region, the slit passed over the transient region during its peak phase. This enabled us to perform a detailed spectroscopic investigation of the transient. We selected \ion{Si}{IV} 1394 \AA\ spectra and fitted with a double Gaussian function to create a spectroscopic intensity image for representation purposes (see left panel of Figure~\ref{fig:spec}). The intensity image clearly shows the transient region, which is marked with a black box. This observation provides a unique opportunity to compare various spectroscopic properties of the transient foot-point at the lower atmospheric height. 

At the transient foot-point, \ion{O}{IV} 1399 and 1401 \AA\ line pair also appeared in the IRIS spectra, which has peak formation temperature $\approx0.14$ MK. We fitted both profiles with Gaussian functions and extracted the intensity counts of both lines (see Figure~\ref{fig:spec}). The intensity ratio of this line pair is sensitive to electron number density as obtained from CHIANTI Version 11 solar soft distribution \citep[see the right panel in Figure~\ref{fig:spec};][]{1997A&AS..125..149D,2024ApJ...974...71D}. Based on their intensity ratio, we obtained the electron number density to be $10^{11.22}$ ($1.66\times10^{11}$) cm$^{-3}$, which corresponds to the chromospheric height. 

IRIS has coverage over several chromospheric and transition region spectral lines such as \ion{O}{IV}, \ion{Si}{IV}, \ion{C}{II}, \ion{Mg}{II}, \ion{S}{I}, \ion{C}{I}, \ion{O}{I}, etc. Many of these lines appear only during the transient activity at the loop foot-point. We fitted average spectral line profiles at the transient location within the maroon box of Figure~\ref{fig:spec} with Gaussian functions and extracted the Doppler velocities and widths. We plot these Doppler velocities and widths obtained from several IRIS spectral lines originating from different ions in Figure~\ref{fig:velocity}. We arranged data points almost in the decreasing order of their spectral line peak formation temperature, which reaches up to the lower atmospheric temperature, and thus probes the lower atmosphere reaching towards the photosphere.  All the spectral lines show redshifted plasma components. Transition region spectral lines such as  \ion{O}{IV} and \ion{Si}{IV} show redshifts of the order of 10-15 km s$^{-1}$, chromospheric line such as \ion{C}{II} shows redshift of the order of 8 km s$^{-1}$ whereas neutral lines such as \ion{C}{I} and \ion{S}{I} show redshift of about 5 km s$^{-1}$. All these redshifted lines indicate that the lower atmosphere is being pushed downward, where the magnitude of velocity decreases as one moves further lower in the atmosphere. As the lower atmosphere is denser compared to the upper atmosphere, this decreasing Doppler velocity may be due to the momentum imparted by the sudden release of energy due to the transient. 

We also found the presence of a triplet of subordinate lines of \ion{Mg}{II} at wavelengths 2791.60, 2798.75, and 2798.82 \AA\  as emission lines (see Figure~\ref{fig:mg2}). Emissions in these lines are very rare and mostly observed in absorption in the solar spectra. Generally, emissions in these lines are observed during a steep rise in temperature in the lower chromosphere more than 1500 K, with electron densities above $10^{11}$ cm$^{-3}$ \citep[see details in ][]{2015ApJ...806...14P}. The peak-to-wing ratio and shape of spectral line profiles of these lines can provide useful diagnostics of the solar lower atmosphere. However, recent modelling studies in flaring conditions and impulsive heating events have shown that the \ion{Mg}{II} triplet lines actually form close in height to the resonance lines in the upper chromosphere, and thus indicate heating at the relatively higher heights \citep[e.g.,][]{2019ApJ...879...19Z,2024MNRAS.527.2523K}.

\subsection{X-ray Analysis}
\label{sec:xray}

During the transient, X-ray emissions were also recorded by XRT/Hinode and XSM/Chandrayaan-2. During the period of observation, only active region AR 12759 was present on the disk in the Sun-Earth line. In the left panel of Figure~\ref{fig:xray}, we plot the associated XRT Al\_poly image of the active region during the transient. We extract the average X-ray flux from the active region within the boxed region and compare its time evolution with the X-ray flux recorded by the XSM (see middle panel of Figure~\ref{fig:xray}). XSM observes the Sun as a star in the energy range of 1-15 keV. It provides a total integrated count within the 1-15 keV energy range at every second, whereas it provides full energy resolved spectra at 1-15 keV at every 60 seconds. We utilized the XSM data analysis software for extracting light curves and spectra \citep[XSMDAS;][]{2021A&C....3400449M}. Both the X-ray light curves show similar flux variation with time, which clearly indicates that the X-ray flux recorded by XSM originated only from the active region present on the solar disk. XSM spectra recorded at the transient time show X-ray flux enhancement between 1.0 and 3.2 keV energy range. Enhanced emission clearly indicates the appearance of heated plasma during the transient event. These spectra also show a few emission lines at 1.3 and 1.8 keV, on top of thermal continuum emission.

We also compare the X-ray flux with the hot emission from the AIA 94 \AA\ passband. In the middle panel of Figure~\ref{fig:xray}, we over-plot integrated emission from the whole active region and full loop obtained from hot AIA 94 \AA\ passband. We scaled the intensities to fit in the plotting range. On comparing all the light curves plotted, we find that X-ray light curves from XSM, XRT, and hot AIA 94 \AA\ emission integrated over the full loop peak at $\Delta t= 44, 55, 54$ s, respectively, whereas hot emission from AIA 94 \AA\ passband integrated over the whole active region peaks later at  $\Delta t=80$ s. All the peak times are noted with respect to the peak observed in the IRIS 1400 \AA\ passband as described in Section~\ref{sec:UVimaging}. This could possibly be due to hot transient emission from other parts of the active region, which are not hot enough to emit in X-rays. We also noticed that transient emission appears first in XSM, followed by XRT, and later in hot AIA 94 \AA\ passband.  We also noted that the emission in AIA 131 \AA\ passband peaks at 18.93 min (see Figure~\ref{fig:lc}), before any other light curves; however, the rise in emission is much faster than the rise in other light curves. This can be interpreted as the appearance of the first very hot plasma due to emission from \ion{Fe}{XXI} 128.75 \AA\ formed around 10 MK \citep{2010A&A...521A..21O,2012SoPh..275...41B}, and then the subsequent appearance of its cooling effect in different light curves. 

\subsection{Energetics of transient}
\label{sec:energetics}

Using the AIA and IRIS observations and diagnostic techniques, we have extracted several plasma parameters that can be utilized to estimate the order of different types of energies released during the transients. Following \citet{2014Sci...346C.315P,2018ApJ...857..137G,2022MNRAS.512.3149G}, the thermal, kinetic, and turbulent energy densities can be written as 

\begin{equation}
 	E_{th}=\frac{3}{2}~n_e~k_B~T; E_{kin}=\frac{1}{2}~n_e~\mu~v^2 ; E_{turb}=\frac{1}{2}~n_e~\mu~ntv^2
\end{equation}

where $n_e$ is electron number density, $k_B$ is the Boltzmann constant, $T$ is temperature, $v$ is flow speed, $ntv$ is non-thermal velocity, and $\mu$ is mean molecular weight ($\approx 0.6m_p$ for fully ionized plasma, where $m_p$ is mass of a proton). Thus, the total energy released in volume $V$ will be given by  $V (E_{th}^h+E_{th}^c+E_{kin}^c+E_{turb}^c) $, where superscripts $h$ and $c$ represent hot coronal and cool transition region plasma, respectively.   

To estimate the energy released during the transient at the foot-point, we use parameters obtained from cool transition region lines (\ion{O}{IV} and \ion{Si}{IV}) during the crossing of IRIS spectroscopic slit at the transient foot-point. We obtained $n_e$ $\approx$ 10$^{11.22}$ cm$^{-3}$, $T$ $\approx 1.4\times10^5$ K (peak formation temperature of \ion{O}{IV} lines), $v$~$\approx$ 10~km~s$^{-1}$. We consider the volume $V$ of the emission to be $\approx 1300~\times1300~\times~1300$ km$^3$, corresponding to a region of 3 AIA pixels cubical structure \citep[e.g.,][]{2014Sci...346C.315P,2022MNRAS.512.3149G}. Using these values, we find that the total energy released at the transition region temperature is $\approx 1.24 \times 10^{25}$~erg. We also estimate the total thermal energy released at the hot temperature using parameters obtained from DEM analysis. We obtained $n_e$ $\approx 3.3\times 10^{9}$ cm$^{-3}$, $T$ $\approx 10.5$ MK, and $V$~$\approx 1300~\times1300~\times~1300$ km$^3$ which results to total thermal energy of the order of  $\approx 1.5 \times 10^{25}$~erg. These estimates suggest the final total energy released to be about  $\approx 2.7\times10^{25}$~erg, including both hot and cool plasma components. These estimates are just the lower limit on energy released during the transient. Estimated energies are equivalent to energies involved in the small micro-flaring events \citep[e.g.,][]{2018ApJ...857..137G,2021MNRAS.507.3936C,2022MNRAS.512.3149G}.

At the loop-top (LT), the obtained peak temperature $T_e$ is about 8.5 MK, and the corresponding density $n_e$ is $\approx 6.2\times 10^{9}$ cm$^{-3}$. This density is higher than the density obtained at the loop foot-point for hot plasma. This is expected as peak intensity obtained from the light curve of AIA 94 \AA\ passband at the loop-top is almost 40\% higher than the corresponding peak intensity at the foot-point.    

\section{Discussion and conclusion}
\label{sec:disc}

During solar flares or transient energy-release events, the solar atmosphere responds very rapidly to the energy deposited in the lower atmosphere. Evolution of the transition region, as well as chromospheric and sometimes photospheric emissions, shows a rapid increase in intensity flux. In this work, we investigated a small-scale transient that occurred at the loop foot-point. The transient was classified as a GOES A-class flare. We studied its evolution in different imaging passbands sensitive to different temperatures, covering the temperature range from 6000 K to more than 10 MK. During the transient, the temperature rose to more than 10 MK, whereas the transient loop attained a temperature of about 8 MK at the loop-top, as estimated from the DEM analysis. We noted the clear time differences between the peak intensity of different passbands sensitive to lower atmospheric emission, whereas not-so-clear time differences were observed between passbands sensitive to coronal emissions at transient foot-points of the flare loop. Time differences between the peak intensity of coronal emissions during the flares are well known and studied, which appear after the emissions in the lower atmosphere are observed \citep[e.g.,][]{2013ApJ...774...14Q,2022MNRAS.512.3149G}. X-ray analysis revealed the appearance of very hot plasma first, and then the subsequent appearance of its cooling effect in different light curves. Sometimes, for some compact transient events such as reported in \citet{2018ApJ...857..137G}, simultaneous peak brightenings are observed in the lower and upper solar atmospheric emissions. In this work, we observed clear time lags between peak intensities for the lower atmospheric emissions, where peak brightening appeared later for cooler temperature plasma. Intensity peak appears first in IRIS 1400 \AA\ at transition region temperature 80 kK, and later in IRIS 1330 at chromospheric temperatures 25 kK, and further in AIA 1700 \AA\ at temperature minimum 4500 K. These time lags clearly show that heat flux is penetrating into the lower atmosphere from the upper atmosphere above. Such a clear time lag response of heating of the lower atmosphere during the small-scale transient is very interesting and has never been reported before. Although the time-dependent upper atmospheric responses at the foot-points of flare loops are well studied and established \citep[e.g.,][]{2012A&A...547A..25P}. Here, heat flux is moving towards the lower atmosphere and reaches up to the photosphere. Beneath the transient, mixed polarity regions were observed where both positive and negative magnetic fluxes were emerging. 

At the transient foot-point, IRIS spectra recorded various heating signatures such as enhanced line width, Doppler shifts of chromospheric and transition region lines, and the appearance of \ion{Mg}{II} subordinate triplet emission lines at wavelengths 2791.60, 2798.75, and 2798.82 \AA. Such Mg II subordinate lines appear very rarely in emission and only when there is a steep change of temperature in the lower chromosphere \citep{2015ApJ...806...14P}. However, recent studies also suggest the formation of \ion{Mg}{II} triplet lines during impulsive heating events can take place close to the upper chromosphere \citep[e.g.,][]{2019ApJ...879...19Z,2024MNRAS.527.2523K}. Therefore, the possibility of transient heating occurring at the relatively higher heights is also possible. Transition region and chromospheric spectral lines such as \ion{O}{IV}, \ion{Si}{IV}, \ion{C}{II}, \ion{Mg}{II}, \ion{S}{I}, \ion{C}{I}, and \ion{O}{I} recorded the downward motion of plasma from the redshift of line profiles with decreasing magnitude in the lower down the atmosphere. These Doppler velocities suggest a down-flowing heated plasma in the lower atmosphere that decelerates in the deeper layers of the lower atmosphere as thickness (or density) increases lower down in the atmosphere. 

The energy released during the flares may be transported from the corona to the chromosphere and further through various transport mechanisms such as via beam of non-thermal electrons \citep[e.g.,][]{1976SoPh...50..153L,2005A&A...435..743S}, thermal conduction \citep[e.g.,][]{2001SoPh..204...91A,2020A&A...644A.172W}, and also downward propagating Alfv\'en waves \citep[e.g.,][]{2024GMS...285...39R}. They all provide different observational signatures in the lower atmosphere.  Heating due to the beam of non-thermal electrons in the small-scale events result in blueshift or upward motion of lower atmospheric ions such as \ion{Si}{IV}, \ion{C}{II}, \ion{Mg}{II}, etc., and the appearance of \ion{Mg}{II} triplet emission lines  \citep{2018ApJ...856..178P}.  However, here, we found redshift or downward motion of these ions, which can be attributed to heating due to thermal conduction \citep{2014Sci...346B.315T,2018ApJ...856..178P}, whereas the appearance of \ion{Mg}{II} triplet emission lines can be attributed to heating due to a beam of non-thermal electrons. Thus, the atmosphere here is getting heated by the hybrid mode of non-thermal electrons and thermal conduction acting simultaneously. In this small-scale event, the continuous emergence of both positive and negative magnetic fluxes next to each other in the mixed polarity region leads to reconnection events and powers the transient. Heating due to this transient event results in either the heat transportation to the lower atmosphere based on the time lags noted in the Section~\ref{sec:UVimaging}, or the local heating of the chromosphere based on the findings noted in \citet{1989SoPh..124..303M,2016ApJ...816...88K,2019ApJ...870..114S} etc. This small transient also powers the heating of plasma to more than 10 MK, which results in enhanced emissions observed at UV, EUV, and X-ray wavelengths. The results presented here will provide useful ingredients for modelling the heat and energy transportation in the solar lower atmosphere during the smaller classes of solar flares. 

\section*{Acknowledgments}
We thank the referee for helpful comments and suggestions that improved the quality of the presentation. The research work at the Physical Research Laboratory (PRL) is funded by the Department of Space, Government of India. AR thanks PRL for her PhD research fellowship. R.E. is grateful to Science and Technology Facilities Council (STFC, grant No. ST/M000826/1) UK, acknowledges NKFIH (OTKA, grant No. K142987 and Excellence Grant, grant No. TKP2021-NKTA-64) Hungary, and PIFI (China, grant number No. 2024PVA0043) for enabling this research. The authors thank Drs. Giulio Del Zanna and Paola Testa for the helpful discussion. AIA and HMI data are courtesy of SDO (NASA). IRIS is a NASA small explorer mission developed and operated by LMSAL with mission operations executed at NASA Ames Research Center and major contributions to downlink communications funded by the Norwegian Space Center (NSC, Norway) through an ESA PRODEX contract.  We acknowledge the use of data from the Solar X-ray Monitor (XSM) onboard the Chandrayaan-2 mission of the Indian Space Research Organisation (ISRO), archived at the Indian Space Science Data Centre (ISSDC). XSM was developed by Physical Research Laboratory (PRL) with support from various ISRO centers. Hinode is a Japanese mission developed and launched by ISAS/JAXA, with NAOJ as a domestic partner, with NASA and STFC (UK) as international partners. It is operated by these agencies in cooperation with ESA and NSC (Norway). CHIANTI is a collaborative project involving George Mason University, University of Michigan (USA), University of Cambridge (UK), and NASA Goddard Space Flight Center (USA).

\section*{Data Availability}

The observational data utilized in this study from AIA and HMI onboard SDO are available at   \url{http://jsoc.stanford.edu/ajax/lookdata.html}, data from the IRIS mission are available at \url{https://www.lmsal.com/hek/hcr?cmd=view-recent-events\&instrument=iris}, data from XRT onboard Hinode are available at \url{https://sdc.uio.no/search/form}, and data from XSM onboard Chandrayaan-2 are available at \url{https://pradan.issdc.gov.in/ch2/}.


\end{document}